\documentstyle[aps]{revtex}

\input BoxedEPS
\SetepsfEPSFSpecial
\HideDisplacementBoxes

\begin{document}
\title{Quantitative measurements of the thermopower of Andreev
interferometers}
\author{D.A. Dikin\footnote{Permanent address: Institute for Low
Temperature Physics and Engineering, Kharkov 61164, UKRAINE.}, S. Jung and V. Chandrasekhar}
\address{Department of Physics and Astronomy, Northwestern University,
Evanston, IL 60208, USA}

\maketitle

\begin{abstract} 
Using a new second derivative technique and thermometers which enable us
to determine the local electron temperature  in a mesoscopic
metallic sample, we have obtained quantitative measurements of the
low temperature field and temperature dependent thermopower of Andreev
interferometers.  As in previous experiments, the thermopower is found to
oscillate as a function of magnetic field.  The temperature dependence of
the thermopower is nonmonotonic, with a minimum at a temperature of
$\simeq0.5$ K.  These results are discussed from the perspective of
Andreev reflection at the normal-metal/superconductor interface.
\end{abstract}


\section{Introduction}
In the last few years, investigations of normal metal/superconductor (NS)
interface structures have shown a wealth of new and interesting phenomena
associated with the penetration of superconducting correlations into the
normal metal.  These effects can be understood as arising from
phase-coherent Andreev reflection of quasiparticles at the NS
interface \cite{Andreev}: an electron is retroreflected as a hole at the NS
interface, with the concurrent generation of a Cooper pair into the
superconductor. The motions of the electron and hole are correlated, and
coupled by the microscopic phase $\phi$ of the superconductor.  The
correlated motion of the electron-hole pairs induced in the proximity
coupled normal-metal wire leads to a modification of its conductance $G$
which depends in a nonmonotonic way on the energy
$E$ of the quasiparticles, with an energy scale set by the correlation
energy $E_c=\hbar D /L^2$  \cite{Thouless} for a diffusive metal
wire of length $L$ and diffusion constant $D$. $G(E)$ has a maximum at
energies $E\simeq E_c$, but approaches the normal state conductance $G_N$
for $E\ll E_c$ or $E\gg E_c$. Experimental evidence for this energy
dependent conductance has been observed in many experiments on the
conductance of NS interface samples as a function of voltage bias or
temperature \cite{Reentrance}. 

For a normal metal connected to two superconductors (a so-called Andreev
interferometer \cite{Takayanagi}), the properties of the normal metal are
sensitive to the phase difference $\Delta \phi=\phi_1 - \phi_2$ between
the superconductors at the two NS interfaces, so long as the electron
phase coherence length $L_\phi\ge L$.  Experimentally, $\Delta \phi$
can be modulated by connecting the two superconductors together and
threading an Aharonov-Bohm $\Phi$ flux through the doubly-connected loop
that results.  The conductance of such an Andreev interferometer will then
oscillate as a function of $\Phi$, with a fundamental period given by the
superconducting flux quantum $\Phi_0=h/2e$.  These conductance
oscillations have been observed in numerous experiments on Andreev
interferometers \cite{AndInt oscill}. 

In addition to the electrical transport properties, the thermal transport
properties of proximity coupled NS structures have also been
investigated \cite{Claughton,Volkov,Heik}.  Experiments on the thermopower
$S_A$ of Andreev interferometers  \cite{Eom,Petrashov} have shown that it
also oscillates as a function of the magnetic flux $\Phi$ coupled through
the interferometer loop.  However, a number of characteristics of the
thermopower of Andreev interferometers are not understood.  First, the
oscillations of $S_A$ can be either symmetric or antisymmetric with
respect to $\Phi$; the symmetry depends on the topology of the
sample, but the reason why this is so is not understood. Second, the
temperature dependence of $S_A$, determined by numerically estimating the
temperature gradient in the sample using a simple heat flow model, was
found to be non-monotonic, with a maximum in
amplitude at a temperature of $\sim$ 140 mK \cite{Eom}, reminiscent of
the reeentrant behavior seen in the temperature dependent resistance of
proximity coupled normal metals.  However, the temperature at which this
maximum occurs is much higher than the characteristic temperature scale
$T = E_c/k_B$ which sets the scale for reentrant behavior, as
demonstrated by measurements of the
temperature dependent
\textit{resistance} of the same samples, which showed no reentrant
behavior in the measured temperature range \cite{Eom}. 

In this Letter, we present new data on the thermopower of
Andreev interferometers, taken using a new second derivative method,
and employing a local thermometry technique \cite{Aumentado} which permits
us to make \textit{quantitative} measurements of the thermopower without
recourse to theoretical modeling.  The Andreev interferometers we have
measured are shorter in length and have a topology different from those in
previous measurements.  Their thermopower is purely antisymmetric in
$\Phi$, as observed before. The temperature dependence of the thermopower
is non-monotonic, but the temperature at which the maximum in amplitude is
observed is in the range of $T\simeq 0.5$ K, indicating that the length
of the sample may play a role in the temperature dependence of
the thermopower. 

\section{Sample fabrication and measurement} 
Figure 1(a) shows a schematic of the sample design.  A dc current $I$ put
through the vertical normal-metal heater wire on the left heats the
electrons in the middle of the heater to a temperature $T_{h}(I)$. 
The thermal voltage across the two contact pads ($V_{th}$) on the
right at the temperature $T_{c}(I)$ has contributions from the Andreev
interferometer and a reference normal metal electrode.  The upper voltage
probe contains the Andreev interferometer (outlined by a dotted line), and
the lower voltage probe contains the reference electrode, which is a pure
gold wire in our case.  The Andreev interferometer probe and the
reference probe join at a single point at the heater, whose temperature
$T_{h}(I)$ can be measured by a proximity effect thermometer (the `hot'
thermometer).  At the other end, both the Andreev interferometer and the
reference wire terminate in a large normal-metal contact whose
temperature can be measured by another proximity thermometer (the `cold'
thermometer).

\begin{figure}[p]
\begin{center}
\BoxedEPSF{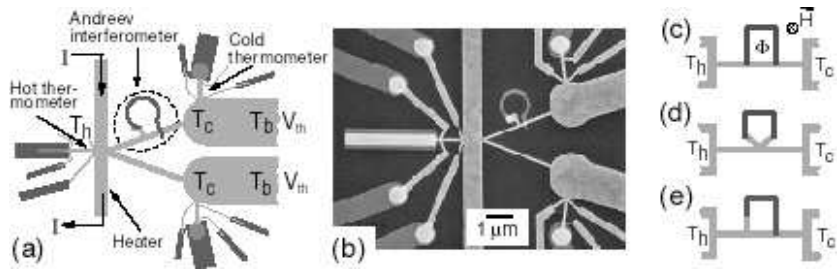 scaled 730}
\end{center}
\caption{
(a) Schematic of our sample design for measuring the thermopower of 
an Andreev interferometer. Dark gray is superconductor, lighter area is
normal metal. The Andreev interferometer is encircled by a dotted line.  
$T_h$ and $T_c$ are the local electronic temperatures at the hot and cold
sides of interferometer, measured by the hot and cold
proximity thermometers. $T_b$ is the base temperature of the  
refrigerator. The thermal voltage is measured using the contacts
labeled $V_{th}$. (b) Scanning electron micrograph of one of our actual
devices fabricated by conventional electron-beam lithography techniques. 
(c), (d) and (e) - schematic of three different types of Andreev
interferometers:  `parallelogram'-, `house'- and `hook'-like
configurations respectively.}
\end{figure}

Figure 1(b) shows a scanning electron micrograph of an experimental
realization of this design implemented using multi-level electron beam
lithography on oxidized silicon substrates.  The 65 nm thick Au normal metal
regions were patterned and evaporated first, after which the 65 nm thick
Al film was evaporated following an O$_2$ plasma etch to ensure
good interfaces between the Au and Al films.  The overall length of
the normal part of the Andreev interferometer and the reference wire is
$\simeq$ 2.7 $\mu$m.  From weak localization measurements on long Au wires
with similar properties, the electron phase coherence length $L_{\phi}$
was found to be $L_{\phi} \sim 3.5$ $\mu$m at $T=300$ mK, and the
diffusion constant in the Au was $D=1.4 \times 10^{-2}$ m$^2$/s, resulting
in a thermal diffusion length $L_T=\sqrt{\hbar D/k_B T}=0.32$ $\mu$m at
$T=1$ K.

In previous work \cite{Eom}, thermopower measurements on
Andreev interferometers of two different topologies were reported.  In
the first (see Fig. 1(c)), denoted the `parallelogram,' a part of the
superconducting part of the Andreev interferometer was placed along the
temperature gradient.  The second (Fig.1 (d)), denoted the `house,'
had no superconducting parts along the temperature gradient.  This
difference gave rise to characteristically different behavior as a
function of magnetic field:  while the thermopower oscillations in the
house interferometer were symmetric with respect to magnetic field, the
thermopower oscillations in the parallelogram interferometer were
antisymmetric with respect to field.  The topology of the samples in this
experiment is similar to the parallelogram geometry, although only one
arm of the superconductor is in the path of the temperature gradient 
(Fig. 1(e)). 

The samples were measured in a $^3$He refrigerator with a base temperature
of 260 mK. The `hot' and `cold' proximity effect thermometers were first
calibrated by measuring their four-terminal resistance with an ac
resistance bridge as a function of the temperature $T_b$ of the
refrigerator, with no current through the heater.  $T_b$ was then kept
constant, and the resistance of the thermometers measured as a function of
the dc heater current.  By cross-correlation of the two measurements, we
could obtain the electron temperatures as a function of heater current
($T_{h}(I)$ and $T_{c}(I)$).  This process was repeated for different
values of $T_b$.  Figure 2 shows the result of this measurement for the
hot thermometer at a few representative temperatures. 

The thermal voltage generated between the two voltage contacts (see
Fig.1(a)) is given by the equation
\begin{equation}
V_{th}=\int_{T_{c}(I)}^{T_{h}(I)}[S_A - S_N] dT.
\label{thermalvoltage}
\end{equation}
Here $S_A$ is the thermopower of the Andreev 
interferometer, and $S_N$ is the thermopower of the reference.  We shall
assume that $S_N$ is small or that it does not vary as a function of
external parameters such as the magnetic field, so that it can be
neglected in our analysis. (Since the reference wire is made from Au, both
conditions are satisfied in our case.)  For small values of the dc
current ($\leq$ 2 $\mu$A), the measured $T_{c}(I)$ is essentially independent 
of temperature and equal to $T_b$\cite{Dikin}.  Taking the derivative of
Eq.(\ref{thermalvoltage}) with respect to current, one then obtains
\begin{equation}
\frac{d V_{th}}{d I}=S_A \frac{dT_h}{dI}.
\label{firstderivative}
\end{equation}
In our previous work \cite{Eom}, the thermopower was determined using this
relation.  $dV_{th}/dI$ was measured by superposing a low frequency
($\sim$ 10 Hz) ac current on top of the dc heater current, and measuring
the resulting ac voltage between two large normal metal contacts kept at
$T_b$ at the same frequency.  By estimating $dT_{h}/dI$ based on a simple
heat flow model, we could then estimate $S_A$.  Since
$T_{h}(I)$ is symmetric in $I$,  $dT_{h}/dI=0$ at $I=0$, so that one
needs to apply a finite dc current in order to obtain a finite ac
voltage.  The application of a finite dc current heats the electron gas,
and complicates the analysis of the temperature dependence.  In order to
avoid this problem, one can consider the derivative
of equation (\ref{firstderivative}) at $I=0$
\begin{equation}
\left.\frac{d^2V_{th}}{dI^2}\right|_{I=0}=S_A
\left.\frac{d^2T_{h}}{dI^2}\right|_{I=0}+\left.\frac{dS_A}{dI}\frac{dT_{h}}{dI}\right|_{I=0}=
S_A\left.\frac{d^2T_{h}}{dI^2}\right|_{I=0}
\label{secondderivative}
\end{equation} 
since $dT_{h}/dI=0$ at $I=0$.  From the measured value
of $d^2V_{th}/dI^2$ and a knowledge of $d^2T_{h}/dI^2$ at $I=0$, one can
determine $S_A$.  $d^2V_{th}/dI^2$ is determined by measuring the ac
voltage at a frequency of $2f$, where the frequency of the ac current
through the heater is $f$ (with no dc current).  $d^2T_{h}/dI^2$ at
$I=0$ is determined by taking the numerical derivative of the curve shown
in Fig. 2.  The resulting values at different $T_b$ temperatures are shown
as an inset to Fig. 2. This procedure gives us a direct \textit{quantitative}
value for $S_A$, without the need to apply the dc current through the
heater and to model the heat flow in the wire in order to estimate the
electron temperature.

\section{Experimental results}
Figure 3(a) shows $dV_{th}/dI$ and $d^2V_{th}/dI^2$ as a function of
the heater current $I_{dc}$ for the Andreev interferometer sample shown in
Fig. 1(a), measured with an ac heater current of rms amplitude 2 $\mu$A.
$dV_{th}/dI$ is antisymmetric and $d^2V_{th}/dI^2$ is symmetric
with respect to $I$, as should be expected for the thermal response, since
$T_{h}(I)$ is symmetric in $I$.  Due to the small signal to noise
ratio in the $d^2V_{th}/dI^2$ measurement, the curve has a small offset. 
To take into account this offset, we match the value of $d^2V_{th}/dI^2$
at $I=0$ with the numerical derivative of the $dV_{th}/dI$ vs $I$ curve
taken at the lowest measurement temperature.

As has been reported before \cite{Eom}, the thermopower of these samples
oscillates as a function of magnetic field, with a fundamental period
corresponding to one superconducting flux quantum $\Phi_0=h/2e$ through
the area of the Andreev interferometer loop. Figure 3(b) shows
$dV_{th}/dI$ at $I_{dc}=$5 $\mu$A as a function of the magnetic field
perpendicular to the plane of the Andreev interferometer (in terms of
magnetic field, $\Phi_0$ corresponds to 22.8 Gauss for this
Andreev interferometer). The thermopower is antisymmetric with respect
to magnetic field. This is similar to the thermoelectric response of the
parallelogram sample of Ref. \cite{Eom}.  For comparison, we also show
in the same figure the oscillations of the resistance of the Andreev
interferometer, which are symmetric with respect to magnetic field.

In order to obtain the temperature dependence of the thermopower $S_A(T)$,
we set the magnetic field at $H=5.7$ G which is equal to $\Phi_0/4$,
corresponding to the maximum value of $dV_{th}/dI$, and measure the
second derivative $d^2V_{th}/dI^2$ as a function of temperature with zero
dc current through the heater.  The filled circles in Fig. 4(a) show
the result of this measuremement, with an ac current of rms amplitude 2
$\mu$A through the heater.  To obtain the thermopower, we must divide
$d^2V_{th}/dI^2$ by $d^2T_{h}/dI^2$ at $I=0$ at the appropriate
temperature.  The circles in the inset to Fig. 2 show the measured
temperature dependence of this quantity.  To obtain values at
intermediate temperatures, we fit the points to a power law in $T$, 
as shown by the dotted line in the inset.  Finally, the thermopower $S_A$ is
obtained by dividing the measured $d^2V_{th}/dI^2$ by these interpolated
values.  The open circles in Fig. 4(a) show the result of this calculation. 
Due to the small value of $d^2T_{h}/dI^2$ at higher temperatures, the data
are very noisy in this regime.  Nevertheless, the temperature dependence is
clearly non-monotonic, with a minimum at
$T\simeq0.5$ K.  As was observed in previous experiments, this temperature
does not appear to be related to the correlation energy
$E_c$, which corresponds to a temperature $T = E_c/k_B \simeq 0.015$ K. 
Evidence for this can be seen in the temperature dependence of the
resistance of the interferometer, which is shown in Fig. 4(b).  The
resistance decreases monotonically below the transition temperature of the
superconductor, showing no hint of reentrance, with an overall change of
about of 3.8$\%$.

\begin{figure}[p]
\begin{center}
\hspace{-1.5cm}
\BoxedEPSF{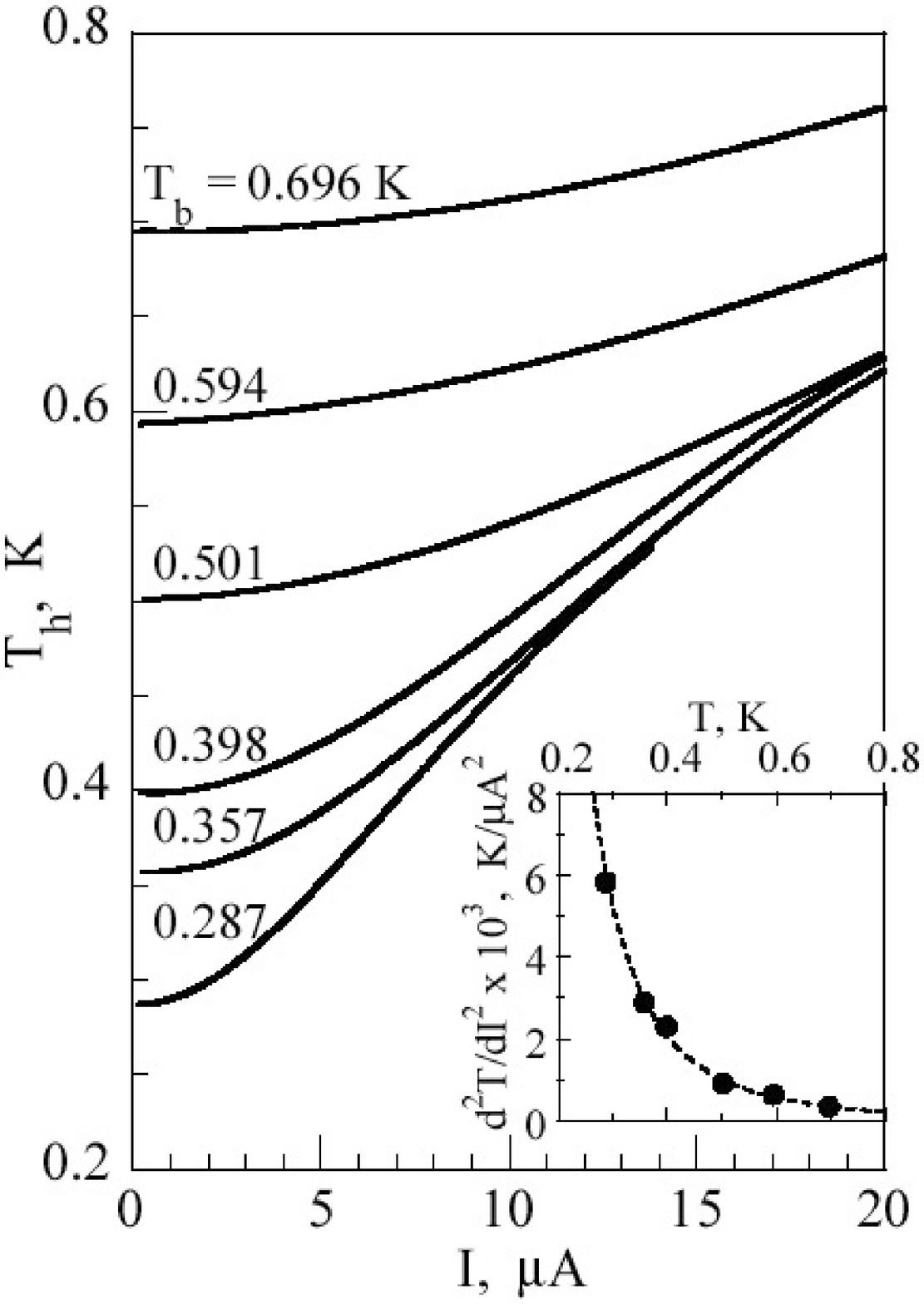 scaled 330}
\hspace{2cm}
\BoxedEPSF{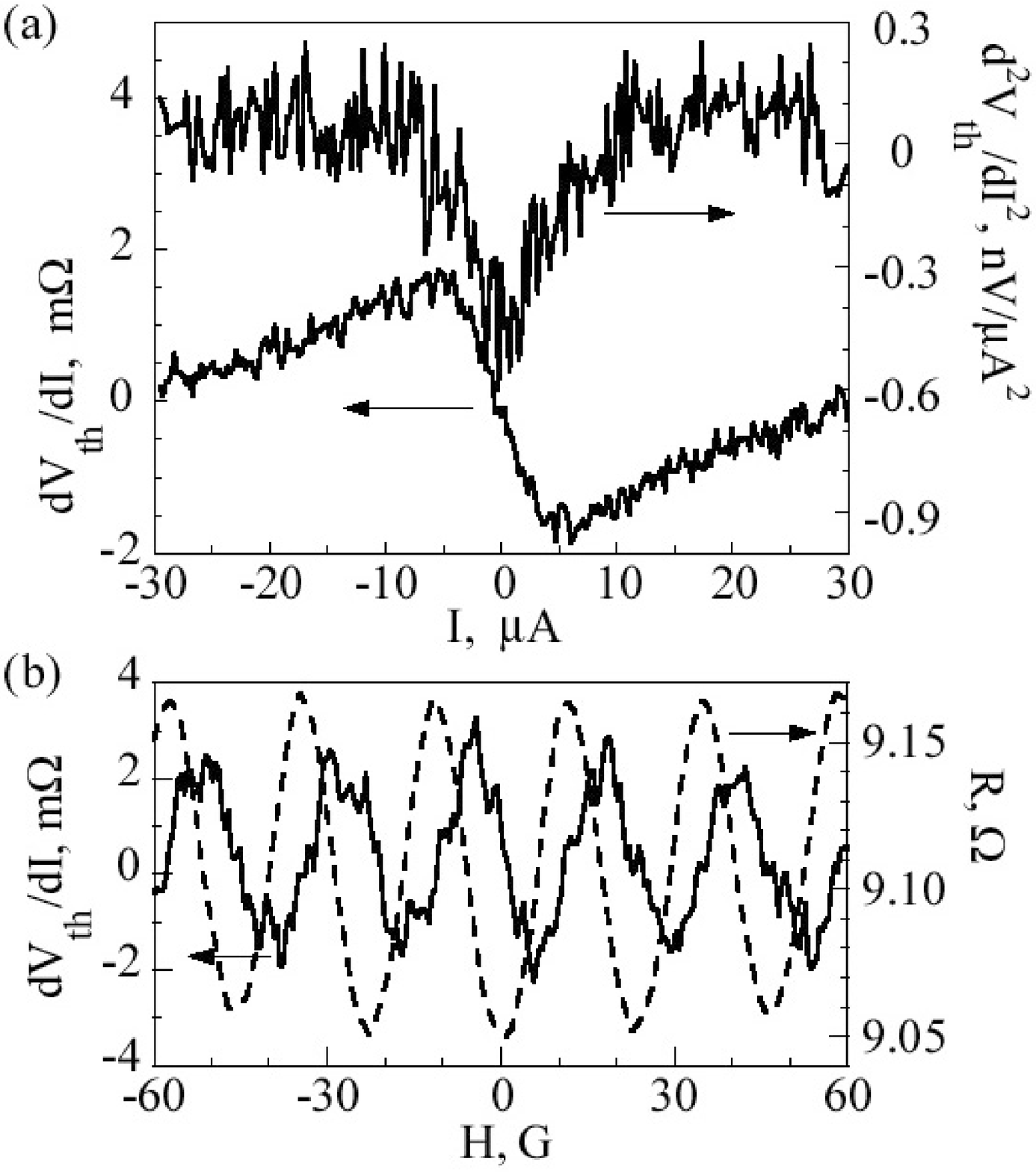 scaled 330}
\end{center}
\caption{Temperature of hot thermometer as a function of dc
current through the heater at different base temperatures $T_b$. Insert
shows temperature dependence of the calculated value of
$d^2T/dI^2$ at $I=0$.  The dashed line in the inset is a fit
to the points by the equation $\frac{d^2T}{dI^2}=1.1\cdot10^{-4}\times
T^{-3.2}$.}
\end{figure}

\begin{figure}[p]
\vspace{-7mm}
\caption{(a) $dV_{th}/dI$ and $d^2V_{th}/dI^2$ as a function of $I$ for
the Andreev interferometer of Fig.1(a). Magnetic field is 5.7 G,
corresponding to a flux 
$\Phi_0/4$ through the area of the interferometer loop. (b) Magnetic field
dependence of the
$dV_{th}/dI$ at $I_{dc}=5 \mu$A (solid line) and resistance of the Andreev
interferometer (dashed line). Dependences were measured at $T_b$ = 0.296K.}
\end{figure}

\vspace{-4mm}
\begin{figure}[p]
\begin{center}
\BoxedEPSF{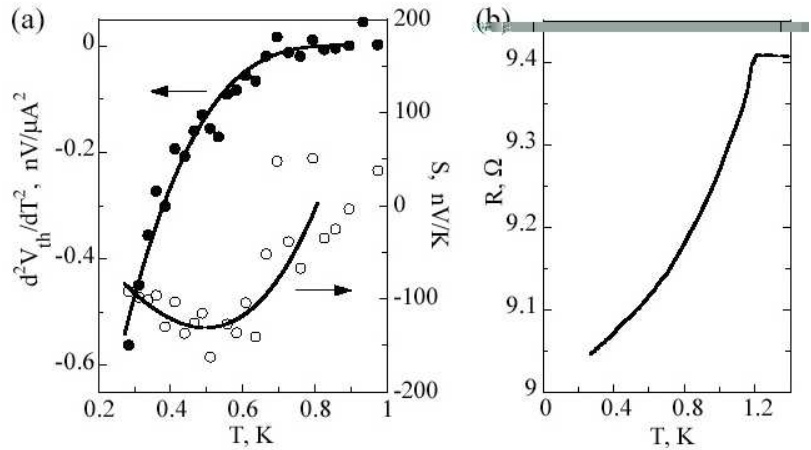 scaled 470}
\end{center}
\caption{(a) Filled circles represent the measured
$d^2V_{th}/dI^2$ as a function of $T$ at a magnetic field of 5.7 G. Open
circles represent the calculated value of $S_A$ as discussed in the
text.  Solid lines are guides to the eye. (b) Resistance of the Andreev
interferometer as a function of temperature.}
\end{figure}

\section{Discussion} 
The most surprising aspect of the thermopower of Andreev interferometers
is the fact that it can be antisymmetric with respect to $\Phi$, even
though the resistance is always symmetric. An antisymmetric magnetic field
dependence implies that the thermopower must change sign as a function
of $\Phi$.  If one associates a negative thermopower with electron-like
quasiparticles, and a positive thermopower with hole-like quasiparticles,
this implies that quantum mechanical interference modulated by the
magnetic field results in a change of the nature of the excitations which
carry the thermal current.

What are the possible origins of this asymmetry?  In numerical
calculations, Claughton and Lambert \cite{Claughton} showed that the
thermopower in mesoscopic NS interferometers could be symmetric or
antisymmetric with respect to field, depending on the topology of the
sample.  However, their arguments depended on the devices having an axis
of perfect mirror symmetry; for samples such as ours without this
feature, the thermopower was predicted to be neither purely symmetric nor
antisymmetric.  A second explanation can perhaps be found in the phase
dependent oscillations of thermoelectric quantities that have been
observed in early experiments \cite{Harlingen} on so-called `thermal
SQUIDs,' which are doubly connected loops in which one arm is fabricated from
one superconductor and the second from another superconductor with a
different gap $\Delta$.  A temperature gradient applied across the loop
will result in a different quasiparticle current in each arm.  This
quasiparticle current is balanced by a counterflowing supercurrent in
each arm of the loop.  Since the quasiparticle currents in each
superconductor are different, the counterflowing supercurrents are also
different, resulting in a net circulating supercurrent in the loop.  The
circulating supercurrent results in a phase gradient which is
proportional to the temperature gradient, and which can be detected by
measuring the critical current of the loop as a function of magnetic
field, for example.  With  this in mind, one might consider the Andreev
interferometer to be a `thermal SQUID,' with one superconductor being the
proximity coupled normal metal arm.  A temperature differential $\Delta
T$ would then result in a phase difference between the two NS interfaces,
and a shift in the thermopower oscillations as a function of magnetic
field. 

A number of experimental observations argue against this
interpretation.    First, the presence of a supercurrent would be expected
to change the phase of the resistance oscillations as well as the
thermopower oscillations.  As we have observed, the thermopower
oscillations are shifted by $\pi/2$, while the resistance
oscillations show no shift with respect to magnetic field.  Second, the
phase shift of the oscillations would depend on the temperature gradient
$\Delta T$, which can be controlled by varying the dc current through the
heater.  However, the phase shift of the thermopower oscillations for this
geometry is always exactly $\pi /2$; no change in the phase of either the
thermopower or the resistance oscillations is observed as a function of dc
heater current.  Consequently, we do not believe a temperature induced
supercurrent is responsible for the asymmetry.  To our knowledge, the
dependence of the symmetry of the thermopower on sample topology has not
been satisfactorily explained. 

The temperature dependence of the thermopower is similar in form to what
has been observed before-- it is non-monotonic, being zero at higher
temperatures, having a maximum in magnitude at some intermediate
temperature, and then going to zero again at lower temperatures.  Although
the electrical transport properties of a variety of NS structures have
been calculated in detail, reliable theoretical predictions for the
thermopower are not yet available. This is because the usual formulation
of the quasiclassical theory of superconductivity which is used as a
starting point for the calculations assumes particle-hole symmetry, and
hence ignores thermoelectric effects from the beginning.  Extensions of
the quasiclassical theory to include thermoelectric effects have proved
difficult\cite{Wilhelm}. Nevertheless, one can make some useful observations
on the differences between the temperature dependence measured here and that
reported in Ref.  \cite{Eom}.  First, the temperature $T_{min}\simeq 0.5$
K at which the minimum in thermopower is observed in this experiment is
greater than the temperature $T_{min}\simeq 0.14$ K observed in Ref.
 \cite{Eom}. The normal line of Andreev interferometer in that experiment
had a length of $\simeq 7 \mu$m; this sample has a length of 2.7
$\mu$m.  For the resistance, $T_{min}$ is expected to scale with
$E_c\sim1/L^2$, i.e., it is larger for smaller $L$, which is the trend we
observe in the thermopower. Second, the overall magnitude of the
thermopower is much smaller than that measured in Ref.  \cite{Eom}, with a
maximum of about 100 nV/K compared to 4 $\mu$V/K observed in Ref.
 \cite{Eom}.  This difference may be understood as arising again from the
topology of the samples.  The Andreev interferometers in this experiment
have a superconductor in the path of the temperature gradient.
Experiments on the thermal conductance of the Andreev interferometers
similar in topology to this sample\cite{Dikin} show that the thermal
conductance of the Andreev interferometer is limited by the thermal
conductance of the small superconducting part.  Hence, most of the
temperature gradient across the interferometer is dropped across the
superconducting part, and very little across the proximity-couple normal
metal, resulting in a small resultant thermoelectric voltage.  This is in
contrast to the `house' interferometer whose temperature dependence was
discussed in Ref. \cite{Eom}, which had no superconductor along the path 
of the temperature gradient.

In conclusion, we have made detailed quantitative measurements of the
thermopower of Andreev interferometers using a new second derivative
technique.  The measurements confirm the two main qualitative observations
from previous experiments: the unusual symmetry of the thermopower with
respect to magnetic field, and the non-monotonic temperature dependence of
the thermopower.  Further theoretical and experimental work is required to
understand the origin of these effects.

\vspace{5mm}
We thank Wolfgang Belzig for valuable discussions.  This work was
supported by the National Science Foundation through DMR-9801892 and by
the David and Lucile Packard Foundation.

\end{document}